# Architectural patterns for handling runtime uncertainty of data-driven models in safety-critical perception

Janek Groß[1], Rasmus Adler[1] [0000-0002-7482-7102], Michael Kläs[1], Jan Reich[1] [0000-0003-1269-8429], Lisa Jöckel[1] and Roman Gansch[2]

[1] Fraunhofer IESE, Kaiserslautern, Germany
[2] Corporate Research, Robert Bosch GmbH, Renningen, Germany

[1] {janek.gross; rasmus.adler; michael.klaes; jan.reich; lisa.joeckel}@iese.fraunhofer.de; [2] roman.gansch@de.bosch.com

**Abstract.** Data-driven models (DDM) based on machine learning and other AI techniques play an important role in the perception of increasingly autonomous systems. Due to the merely implicit definition of their behavior mainly based on the data used for training, DDM outputs are subject to uncertainty. This poses a challenge with respect to the realization of safety-critical perception tasks by means of DDMs. A promising approach to tackling this challenge is to estimate the uncertainty in the current situation during operation and adapt the system behavior accordingly. In previous work, we focused on runtime estimation of uncertainty and discussed approaches for handling uncertainty estimations. In this paper, we present additional architectural patterns for handling uncertainty. Furthermore, we evaluate the four patterns qualitatively and quantitatively with respect to safety and performance gains. For the quantitative evaluation, we consider a distance controller for vehicle platooning where performance gains are measured by considering how much the distance can be reduced in different operational situations. We conclude that the consideration of context information of the driving situation makes it possible to accept more or less uncertainty depending on the inherent risk of the situation, which results in performance gains.

## 1 Introduction

Data-driven models (DDM) based on machine learning are an enabler for many innovations. A huge field of application concerns the perception of the environment in increasingly autonomous systems. Considering self-driving road vehicles, DDMs can be used, for instance, to detect and classify traffic participants or road signs. The main issue that limits the usage of DDMs for such perception tasks is the assurance of safety. A solution that would make it possible to use DDMs for realizing safety-critical functionalities would be extremely valuable for all industries dealing with safety.

A major safety concern is that DDM outputs are subject to uncertainty due to the inherent statistical nature and implicit definition of their behavior, which are mainly based on the available training data. One promising approach to address this safety concern is to estimate the uncertainty for a particular output during operation and to handle it by adapting the system behavior. This approach leads to two related challenges. The first one is how to determine uncertainty during operation in a dependent way. The second one is how to handle estimated uncertainties by means of behavior adaptations. Most existing work focuses on the first challenge. In previous work, we also already proposed and evaluated a solution for estimating uncertainties during operation [1, 2, 3, 4]. Therefore, this paper focuses on the second challenge, i.e., uncertainty handling and the interface to uncertainty estimation. In a previous paper, we elaborated one possible solution for handling uncertainties [5]. In this paper, we systematically derive alternative patterns and evaluate them by means of a simple application example.

As for the application, we will consider a distance controller for keeping a safe distance between two vehicles. Traditionally, such a distance controller is realized without any DDM. Thus, we use this traditional approach as a reference baseline and evaluate how much we could reduce the distance if we use a DDM to consider additional information; that is, the friction coefficient of the leading vehicle. Traditionally, a fixed worst-case value would be defined for this friction coefficient. We hypothesize that using a dynamically estimated value instead of the worst-case value bears utility potential. However, using a DDM comes with uncertainties in the predicted friction value. Therefore, the achievable utility gain varies depending on the level of uncertainty that we can realistically achieve and accept, but also on the architectural pattern we apply to handle uncertainty at runtime. This raises the research questions we address in this work: ***RQ1*** – What are general patterns for dealing with DDM-related runtime uncertainty on an architectural level? ***RQ2*** – Which advantages and disadvantages does each pattern have compared to (a) using worst-case approximation and (b) situation-independent uncertainty estimates obtained at design time? ***RQ3*** – How do relevant parameters such as the prediction performance of the DDM and the accepted level of uncertainty affect the perceived utility gain?

This paper offers the following contributions to answer these questions: (1) systematically derived architectural patterns to deal with DDM-related uncertainty and (2) an initial evaluation of these patterns. The evaluation comprises (a) an implementation of the distance controller example; (b) an analysis of our implementation by means of an evaluation of the impact of parameters such as the degree of accepted uncertainty; and (c) a comparison and discussion of the different patterns.

The paper is structured as follows: First, we will discuss related work and provide a quantitative definition of uncertainty. Second, we will introduce the running example. Third, we will answer *RQ1* and describe the architectural patterns for handling uncertainty. Fourth, we will present our approach for evaluating the patterns by means of the example. Afterwards, we will present the evaluation results and use them to answer the *RQ2* and *RQ3* before concluding the paper.

## 2 Related Work

Our proposed patterns for handling uncertainty estimates at runtime are related to approaches that can provide these uncertainty estimates. They complement each other like error detection and error handling. Our work is more closely related to approaches that estimate uncertainties during operation for a concrete DDM output. However, there is also a link to approaches that estimate a situation-independent general uncertainty value at design time, e.g., by statistical testing, because the patterns are also applicable if the uncertainty is assumed constant. Uncertainty estimates during operation can be determined, e.g., by extending existing models like Bayesian neural networks or deep ensembles [6], or by using model-agnostic approaches like uncertainty wrappers [1]. The latter considers a required confidence level in the uncertainty estimates, which is preferable for safety-related contexts.

A basis for combining approaches for uncertainty estimation with approaches for uncertainty handling is a clear *interface* with an unambiguous definition of uncertainty. In previous work, we related uncertainty to the probability that DDM output or a statement about the outcome is not correct [1]. More formally, *uncertainty* is the complement of certainty, where certainty is a lower bound on the probability of correctness justifiable on a given level of confidence considering the current state of knowledge. In the following, we will use the term *data-driven component* (DDC) for a component that comprises a DDM but enhances the DDM output with uncertainty information.

In this paper, we relate uncertainty to a certain safety-critical failure mode like too high or too low for a given output of a DDM. Our estimated uncertainty is thus related to the probability of a failure mode. The difference is that 100% uncertainty means that we do not know at all whether the failure mode is present, whereas 100% failure probability means that the failure mode is definitely there.

Our patterns for handling uncertainty abstract from the causes that contribute to uncertainty. Their application is thus not limited to DDCs and also relates to the handling of random events in the environment or other kinds of random events considered in existing safety standards such as IEC 61508. However, following traditional safety standards, it is not common to estimate failure probabilities during operation. Safety analyses such as fault tree analyses are used at design time to identify the relationship between causes and top-level system failures and perform related probabilistic reasoning. For instance, the occurrence probability of every cause is estimated and the occurrence probability of the top-level failure is derived. However, in practice, all this is done at design time. The concept of component fault trees [7] supports automation of such analyses by modularizing fault trees and making them composable so that compositional fault trees can be generated and analyzed when components are composed into systems at design time. In the context of the DEIS project [8], related runtime analyses concepts were developed, but this work did not consider DDCs. Accordingly, runtime estimation of uncertainty was not considered and all probabilities of causal events were statically defined at design time.

As mentioned above, uncertainty estimation and uncertainty handling complement each other like error detection and error handling. In this sense, the patterns for handling uncertainty are related to error handling. This can be seen as a special case of uncertainty handling that considers only the uncertainty values 0 and 1. Reactions concerning values in between are not considered. We address this gap with our approach.

Salay et al. propose to work with an imprecise world model comprising a set of precise world models to handle uncertainties in perception, yet they do not further elaborate their idea by proposing an architectural pattern that would be a reference to implement their idea, neither do they investigate feasibility of assumptions or implications on performance [9]. Henne et al. proposed an architectural pattern for handling uncertainties at different stages of the perception chain [10]. These uncertainties are fed into a dynamic dependability management component that merges outputs from the perception chain and from a verified low-performance safety path. Compared to our patterns, this pattern is at a much higher level of abstraction. For instance, it does not describe which uncertainty information is delivered and how it is processed.

The Situation-Aware Dynamic Risk Assessment (SINADRA) approach uses Bayesian networks to determine the likelihood that a possible system behavior in the current operational situation will lead to an accident [11]. This approach is related to some patterns that allow dynamic adaptation of a threshold for acceptable uncertainty and can be used as an alternative solution to our patterns. The estimated uncertainties could be fed into Bayesian networks, so the accident likelihood would consider not only uncertainties due to the behavior of other traffic participants and other environmental aspects but also the uncertainty that environmental aspects might not be perceived correctly.

As our patterns are not limited to the handling of uncertainties of DDCs, many approaches for estimating perception uncertainties could be compatible with our patterns. However, most approaches focus on design-time estimation of uncertainty. For instance, the work in [12] presents an approach for expressing perception uncertainties of LIDAR by means of Bayesian networks.

## 3   Example use case

As an example for explaining and evaluating our patterns, we will consider a vehicle function intended to ensure a safe distance to the vehicle in front. We consider the safe distance rule from [13], which formalizes the safe distance as follows:

$$d_{safe} = \left[ v_F \rho + \frac{1}{2} a_{max,acc,F}\, \rho^2 + \frac{(v_F + \rho\, a_{max,acc,F})^2}{2 a_{min,brake,F}} - \frac{v_L^2}{2 a_{max,brake,L}} \right]_+ \quad (1)$$

The first three terms together represent the stopping distance of the follower vehicle considering (i) the reaction distance based on follower speed $v_F$ and reaction time $\rho$ that is required until the follower can initiate the braking maneuver after the leader has started to brake, (ii) the acceleration distance (assuming the follower constantly accelerates with $a_{max,acc,F}$ during reaction time), and (iii) the follower braking distance when the follower constantly brakes with deceleration $a_{min,brake,F}$. The last term represents the leader's braking distance. To estimate the safe distance $d_{safe}$, we subtract the leader's braking distance from the follower's stopping distance. Using this formula, we can formalize the safety constraint "keep a safe distance" with $d_{current} \geq d_{safe}$ where $d_{current}$ refers to the current distance. To discuss the handling of uncertainties, we focus on the road friction coefficient of the leading vehicle $\mu_L$, which affects the lead vehicle's traction during braking. We assume that the leader's road friction is estimated by means of a DDC, e.g., one of those used in the context of [14] or [15]. The following two equations concretize the physical relationship between $a_{max,brake,L}$ from the safe distance formula and the friction coefficient $\mu_L$.

$$a_{max,brake,L} = min\left(\frac{F_{b,traction,limit,L}}{m_L}, \frac{F_{b,brakesystem,limit,L}}{m_L}\right) \quad (2)$$

$$F_{b,traction,limit} = m_L \cdot g \cdot \mu_L \quad (3)$$

In general, the effective maximum leader deceleration limit $a_{max,brake,L}$ is influenced by all driving resistance forces acting in the longitudinal direction (brake system brake force, air resistance, rolling resistance, road inclination resistance). For the sake of exemplification, Equation (2) only considers the effective deceleration to be bound by the maximum brake force the brake system is capable of generating at the wheels and the traction limit determining how much of the generated force can be transferred effectively to the road. While the brake system's limit is mainly influenced by construction, brake pad wear, and the pedal force a driver is likely to apply, the traction limit depends on the vehicle's mass $m$, the gravitational constant $g$, and the friction coefficient $\mu$. We assume that the friction coefficient $\mu_L$ of the leading vehicle is estimated by a DDM and thus subject to uncertainty. In previous work [5], we argued that it is reasonable to use a DDM, apart from the safety challenge that we address with our approach.

In order to apply our architectural patterns, we consider the architecture depicted in Figure 1, which shall ensure the safety constraint $d_{current} \geq d_{safe}$, as a starting point.

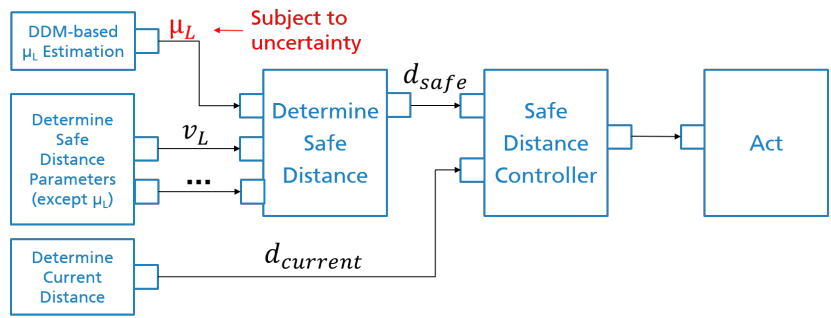

**Fig. 1.** Architectural perspective on the function ensuring the safe distance constraint

A 'safe distance controller' receives the current distance and the safe distance as input. It checks whether the current distance approaches the safe distance and sends brake commands so that the current distance does not go below the safe distance. For this purpose, the RSS framework [13] defines the concept of 'proper response' to ensure that the safety distance constraint is met.

The current distance is measured directly, whereas the safe distance is derived according to the formulas mentioned above. High-integrity distance measurements are possible with radar sensors and are already used in series, e.g., in adaptive cruise control systems. Accordingly, one component determines the necessary values of variables like μ_L and another component calculates the safe distance from these values.

The critical failure mode of the estimated friction coefficient μ_L of the leading vehicle is 'too low' because the braking capability of the leading vehicle would be underestimated, causing that the safe distance to the leading vehicle would to be too small. Accordingly, we focus on the uncertainty with respect to this critical failure mode.

## 4   Architectural patterns for dealing with uncertainty

In this section, we address *RQ1* 'What are general patterns for dealing with DDM-related runtime uncertainty on an architectural level?' by proposing and discussing patterns for handling uncertainty. We consider the robotic paradigm *sense-plan-act* and focus on the uncertainty that a safety-critical failure mode in the sensing or perception of the environment is present. In our example case, the uncertainty that the determined friction coefficient $\mu_L$ of the leading vehicle has the failure mode 'too low' causes an uncertainty regarding the fulfillment of the safety constraint. The objective of all patterns is to limit this uncertainty to an acceptable level. For the remainder of this work we will focus on $\mu_L$ as the uncertainty afflicted variable and assume that other variables are either measured with perfect accuracy or are worst-case estimates. We only consider established platoons and disregard situations where a platoon is formed or dissolved.

Next, we will first systematically derive the patterns and then briefly discuss implications if we need to deal with multiple uncertain variables.

We see two elementary *design decisions* for the patterns. The first one addresses the interface, i.e., whether we provide a single value $x$ and related uncertainty $u$ or an uncertainty distribution regarding possible values $u(x)$. The second decision addresses whether we assume a fixed target for the degree of acceptable uncertainty $u_{acceptable}$ or consider the option that $u_{acceptable}$ depends on further situational information $y$.

In our further discussion, we will focus on the resulting alternative *uncertainty handlers*, which are illustrated in Figure 2, and which would be placed between the sensing component providing the required information and the decision component using it. In our example architecture given in Figure 1, it would be placed between the DDC providing $\mu_L$ and the component determining the safe distance.

The **uncertainty supervisor** shown in the top-left part of Figure 2 receives an input $x$ that is estimated by a DDC. Furthermore, it receives an uncertainty $u$ representing the uncertainty that $x$ has a certain failure mode. It compares this uncertainty $u$ with a fixed threshold $u_{acceptable}$ defining what is acceptable from a safety perspective. If the uncertainty is not acceptable, it overwrites the input variable $x$ with a default value, e.g., the worst-case value. In our example, the variable $x$ would be the friction coefficient $\mu_L$ and the uncertainty $u$ would refer to the failure mode 'too low'. As default value, we would choose a value where we can assure that it is the highest value ($\mu_{max}$) that the leading vehicle could encounter in the intended usage context.

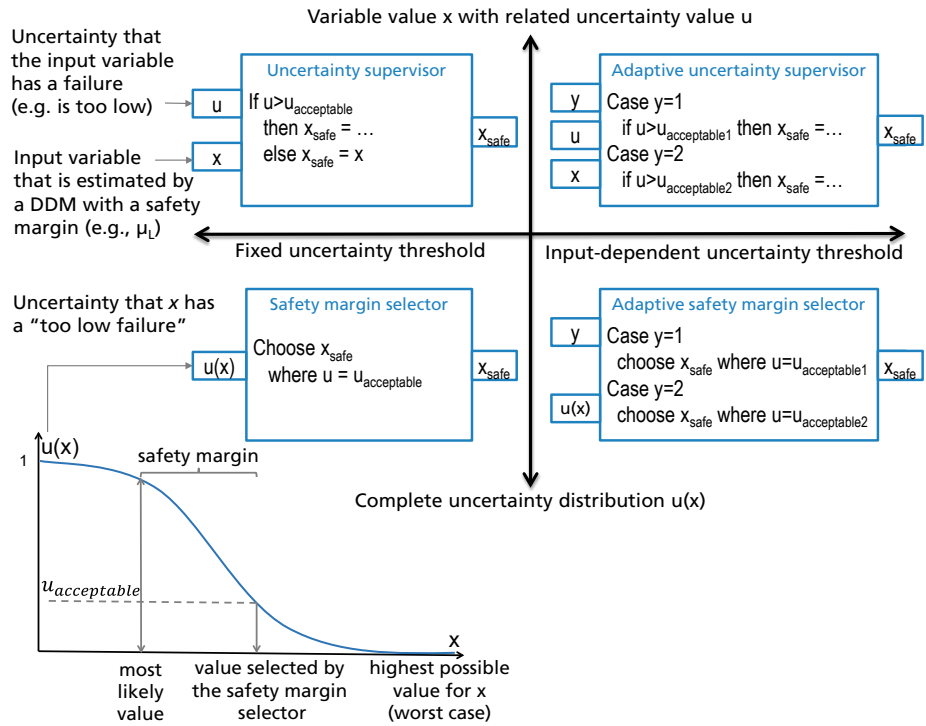

**Fig. 2.** Checker options to implement the four architectural patterns for handling uncertainty

The *adaptive uncertainty supervisor* in the top-right part of Figure 2 has an additional input $y$ that is used to adapt the acceptable uncertainty threshold. This can make sense because the safety criticality of the DDC output may depend on the current operational situation. In our example of assuring a safe distance, a rear-end collision would be less severe if the vehicle speed is very low. Accordingly, one could think of two safety goals addressing different situations with different integrity levels and $u_{acceptable}$. If we apply this approach, we need to consider that situations must not be too fine-grained as described in clause 6.4.2.7, part 3 of ISO 26262: 'It shall be ensured that the chosen level of detail of the list of operational situations does not lead to an inappropriate lowering of the ASIL [of the corresponding safety goals].'

The *safety margin selector* in the lower-left part of Figure 2 receives an uncertainty distribution, which is also illustrated in the figure. The x-axis refers to value $x$, e.g., the friction coefficient $\mu_L$. The y-axis assigns each x-value the associated uncertainty $u(x)$. In our example, $u(x)$ would be the uncertainty that the respective $\mu_L$ value is 'too low'. This implies that higher $x$ values have lower uncertainties and the highest practically possible $x$ value $\mu_{max}$ has uncertainty zero. Based on this distribution, the safety margin selectors choose the $x$ in a way that $u(x) \leq u_{acceptable}$.

The *adaptive safety margin selector* in the lower-right part finally combines an uncertainty distribution with an adaptable uncertainty threshold.

So far, we have assumed that only the friction coefficient of the leading vehicle is estimated by a DDM. Yet, we can apply the patterns in the same way for further uncertain variables when budgeting uncertainty to them. In consequence, the threshold for the friction coefficient would be lower and an uncertainty supervisor would overwrite the estimated friction coefficient more frequently with the worst-case default value. Similarly, a safety margin selector would choose more pessimistic friction values.

Considering uncertainties from many variables, (static) budgeting of the uncertainty threshold would lead to behavior that is not optimal in all situations. If we consider uncertainty supervisors, in some situations some variables might not need the uncertainty budget that has been assigned to them, while other variables would be overwritten as they are just above the uncertainty threshold. If we consider safety margin selectors, we may have a similar effect, as the safety margins are not selected optimally. An approach to overcoming this issue is to propagate the uncertainties and then apply an uncertainty handler after all uncertainties are combined. In our example, an uncertainty handler for $d_{safe}$ would replace uncertainty handler for $\mu_L$ and further input variables.

However, this requires that the 'determine safe distance' component appropriately integrates the different variables and their uncertainties in the calculation of $d_{safe}$.

## 5  Simulation-based evaluation approach

This section we describe how we simulated the application of the previously derived patterns to answer *RQ 2* and *RQ 3* in Section 6. To answer our RQs in the example use-case, we need to evaluate the impact that each pattern has on the safe distance $d_{safe}$ when we estimate the leader's friction coefficient $\mu_L$ by means of a DDC and use the patterns to deal with related uncertainties. For this purpose, we need to generate the parameter values from which we can calculate $d_{safe}$. In the following, we will thus first provide an overview of our approach to generating this information. Then we will describe the assumptions we made, i.e., the values of the constants and the anticipated distributions. Finally, we will report technical aspects of our implementation.

*Overview* – Figure 3 illustrates the data flow of the implemented simulation approach. As shown in the left part, we use two random variables called *Weather Conditions* and *Behavioral Conditions* as input to create a variation in the considered situations. The *Weather Conditions* are used to generate situational friction information and related uncertainty information as we assume that the friction depends on the situational weather conditions. We vary the estimated friction value μ as well as the related dispersion $\sigma$ to get a situational probability distribution. We define a situational uncertainty threshold $u_{acceptable}$ dependent on the behavioral condition that a human is actively supervising the vehicle because we assume that supervision leads to less strict uncertainty thresholds. The generated information is used to apply the different patterns and calculate their output, namely the friction coefficient $\mu_{safe}$. As illustrated in Figure 3, we also use the *Behavioral Conditions* to generate velocity values for both vehicles. These velocity values are used together with μ_safe and the predefined *Safe Distance Parameters* to compute the minimum required safe distance $d_{safe}$.

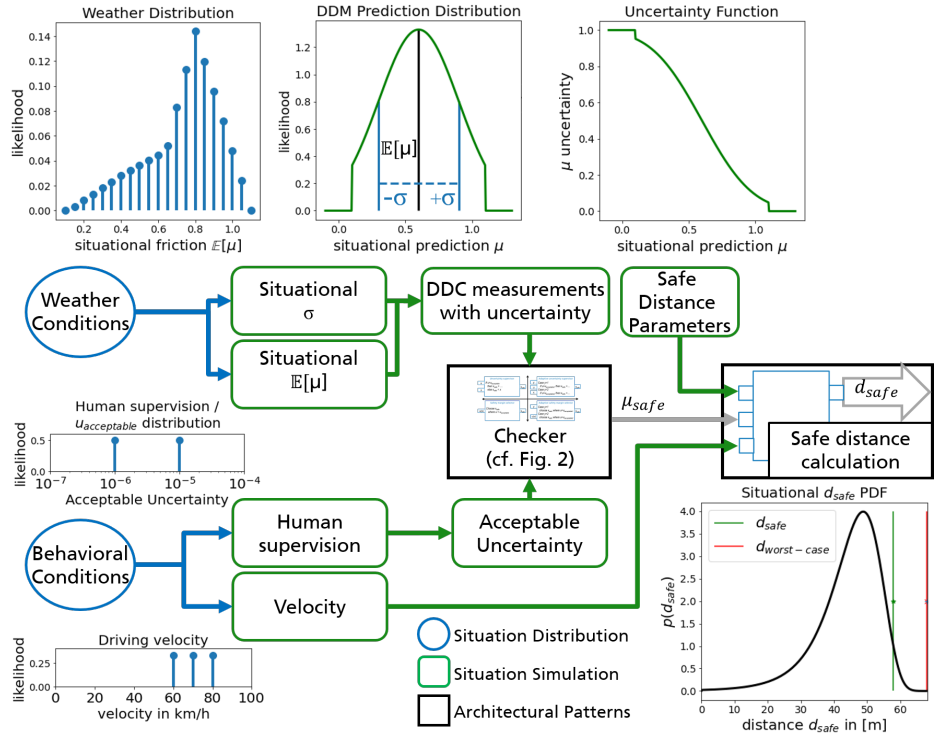

**Fig. 3.** Overview of the implementation approach for answering the research questions

*Assumptions* – Discussing the underlying simulation assumptions, we start from the bottom left in Figure 3 where we defined several distributions that characterize the simulated situations. To generate the vehicle velocities from the *Behavioral Conditions*, we assumed a discrete distribution of velocities between 60 and 80 km/h in order to comply with speed limits for commercial vehicles on German highways. We considered only situations with established platoons where both vehicles drive at the same speed. For specifying the acceptable uncertainty, we assumed a distribution where the driver supervises the distance controller 50% of the operating time and $u_{acceptable} = 10^{-5}$ in the case of supervision and $10^{-6}$ in the case of no supervision.

The *Weather Conditions* are defined as a discrete unidimensional distribution. This distribution describes multiple degrees of one of the four conditions (i) dry weather, (ii) light rain, (iii) snow, and (iv) heavy rain/freezing rain with a respective frequency of 300, 100, 60, and 5 days per year. For the distribution of *Situational Friction* coefficients, we considered empirical friction measurements [16] and assumed the road conditions dry asphalt, wet asphalt, snow/wet leaves, glaze/aquaplaning with the respective friction coefficients 0.8, 0.64, 0.41, and 0.14. To achieve a more fine-grained distribution, we used friction values in steps of 0.05 and derived the corresponding likelihoods from the *Weather Conditions* distribution through linear interpolation.

To model the inaccuracy of the DDM-based friction predictions, we assumed unobservable random variables including events like sensor noise or road surface irregularities leading to disparities. For simplicity, we assumed independent unobservable random variables that influence the DDM friction prediction uncertainty. According to the central limit theorem, the distribution of the mean of *n* independent and integrable random variables converges to the normal distribution with increasing *n*. Thus, we assumed that the friction coefficients are normally distributed with expected situational friction E[μ] as mean and situational dispersion $\sigma$ as standard deviation. However, both the maximal and minimal friction on the road are limited by physical constraints. We therefore assumed μ = 0.1 as the lower limit and μ = 1.1 as the upper limit and cut off the normal distribution at these limits. The dispersion $\sigma$ is assumed to decrease linearly with better weather conditions and higher friction values with a maximum dispersion of $\sigma = 0.075$ for glaze and a minimum of $\sigma = 0.02$ under perfect road conditions.

Further assumptions relate to the simulated DDC and the architectural patterns themselves. For the two uncertainty supervisors, the DDC returns a tuple of the prediction and a corresponding uncertainty value $(\mu_L, u)$. The uncertainty is computed using the cumulative distribution function $F$ of the situation-dependent friction distribution where $u = P(\mu > \mu_L) = 1 - F(\mu_L)$. We considered a safety margin Δμ with $\mu_L = \mu_{predicted} + \Delta\mu$ and the determined by a grid search optimization on the safe distance. Thereby, we chose the Δμ value that minimizes the expected safe distance over all situations. For the safety margin selector patterns, the DDC returns the inverse of the complete uncertainty function $(1 - F)^{-1}$.

For the *Safe Distance Parameters* that we considered as constants, we assumed a maximum acceleration of the following vehicle of $acc_f = 2\frac{m}{s^2}$ and the gravity acceleration of g = 9.81$\frac{m}{s^2}$. To simulate both a highly efficient and a human-like reaction time, we ran our simulation with either ρ = 0.1s for use case A or ρ = 0.8s for use case B.

*Implementation* – We used the programming language Python3 [17] and Jupyter Notebooks [18] in a web-based development environment supported by interactive widgets using the ipywidgets library. Using these widgets, interactive plots were created in combination with the matplotlib [19] library for rapid prototyping. The implemented probability distributions are specified using NumPy [20] and SciPy [21], two libraries for linear algebra, statistics, and scientific computing in Python.

## 6     Study Results and Discussion

In this section, we will present and discuss the results of our simulation study based on the study design and implementation presented in the previous section.

*RQ2 – Comparison of patterns with static approaches* – To estimate the expected utility gain from dynamic handling of uncertainty and to compare the patterns for two anticipated use cases A and B, the expected utility, measured as the expected reduction in the required safe distance $d_{safe}$, was determined using the assumed distributions over the situational and behavioral distributions as introduced above. The observed expected friction $E(\mu_L)$ and distances $E(d_{safe})$ are summarized in Table 1.

**Table 1.** Comparison between different patterns for uncertainty handling of estimated friction values. The columns contain safe distance calculations for use case A - platooning with a low latency distance controller and use case B - platooning with human reaction time. Distances are compared between scenarios that use (a) the worst-case assumption, (b) static design time uncertainty estimates and (c) the four proposed patterns for dynamic uncertainty handling at runtime. Each cell contains the expected minimum required safe distance and the corresponding expected minimum safe friction that is assumed for the leading vehicle (lower is better).

| Expected distance [m] (expected friction [1]) | Use Case A – Platooning reaction time $\rho$ = 100 ms | | Use Case B – Default reaction time $\rho$ = 800 ms | |
|---|---|---|---|---|
| (a) Worst-case $\mu_{max}$ | 14.670 m (1.100) | | 33.350 m (1.100) | |
| (b) DDC with static uncertainty estimate | 13.727 m (1.060) | | 32.407 m (1.060) | |
| (c) DDC with dynamic uncertainty estimates | Single value $(\mu_L, u)$ | Distribution $u(\mu_L)$ | Single value $(\mu_L, u)$ | Distribution $u(\mu_L)$ |
| Constant threshold $u_{acceptable}$ | 12.649 m (1.011) | 10.638 m (0.922) | 31.329 m (1.011) | 29.318 m (0.922) |
| Input-dependent $u_{acceptable}(y)$ | 12.305 m (1.000) | 10.364 m (0.913) | 30.985 m (1.000) | 29.044 m (0.913) |

*Interpretation:* As expected, the worst-case baseline performed worse than all other patterns in terms of utility in both use cases. Considering uncertainty at design time with a situation-independent uncertainty estimate provides some benefits compared to this baseline but is outperformed by any of the dynamic situation-aware uncertainty patterns. For the dynamic patterns, the use of an input-dependent acceptable uncertainty threshold $u_{acceptable}$ led to utility improvements as well as to propagating uncertainty as a distribution $u(\mu_L)$ instead of a single value.

The magnitude of improvement differs depending on the use case. For use case A, which focuses on platooning with a highly efficient solution with short reaction times, the best solution reduced the average requirement on $d_{safe}$ by ~29%. In use case B, which focuses on default driver assistance with a more relaxed requirement on the reaction time, the reduction was only ~13%, which could increase the capacity on the road under optimal conditions by approximate the same amount.

*RQ3 – Sensitivity analysis of the results –* To understand the impact on key parameters of the perceived simulation outcomes, we conducted a sensitivity analysis. Sensitivity analysis hereby refers to an analysis that is applied to determine how an output variable is affected by changes in one or more input variables. We investigated the effects of choosing different thresholds for the accepted uncertainty $u_{acceptable}$ and varied the dispersion σ of the predicted friction values $\mu_L$ provided by the DDC. This corresponds to using a DDM in the DDC that is either more or less accurate in its predictions.

Figure 4 provides the results of the sensitivity analysis with respect to the acceptable uncertainty $u_{acceptable}$ considering situations that differ regarding the friction $\mu_L$.

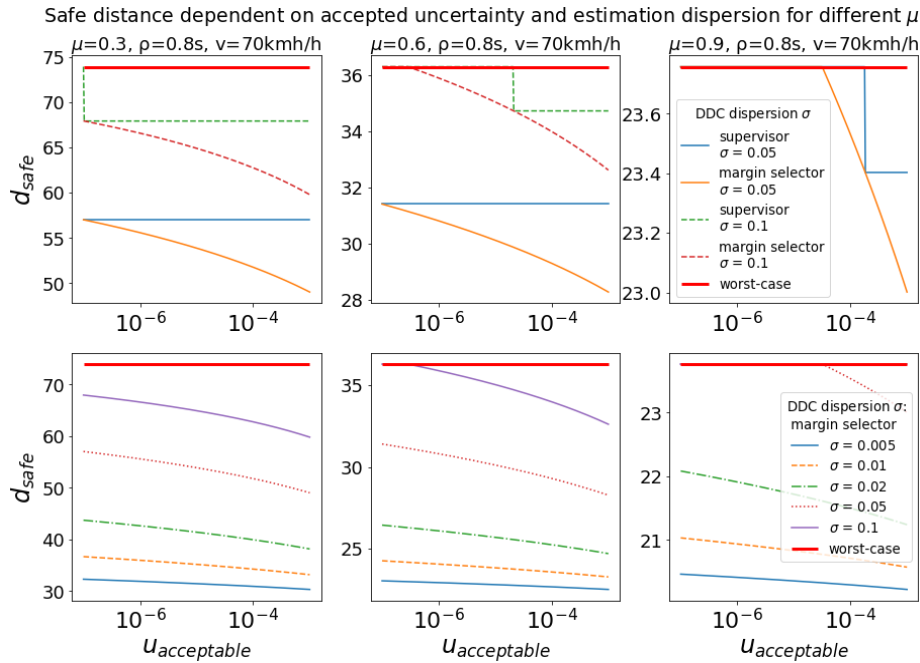

**Fig. 4.** Sensitivity of the safe distance to the accepted uncertainty threshold for multiple levels of dispersion $\sigma$. In the first row the proposed margin selector and uncertainty supervisor patterns are compared. In the second row, the safe distance is computed only using the margin selector.

*Interpretation:* The results in the first row of Figure 4 illustrate that margin selectors are more flexible compared to supervisors in dealing with different thresholds on the accepted uncertainty. In the case of $\mu = 0.9$ and $\sigma = 0.1$, the patterns do not yield any benefit over the worst-case assumption. The results in the second row of Figure 4 indicate that the safe distance requirement is less sensitive to DDCs that have a low accuracy, i.e., a high dispersion $\sigma$. For stricter, i.e., lower, acceptable uncertainty thresholds, such DDCs do not outperform the worst-case baseline, yet they are more sensitive to changes in the threshold on the accepted uncertainty. On the other hand, for very accurate DDCs, the safe distance requirement is significantly reduced and the reduction is hardly affected by the acceptable uncertainty threshold.

In conclusion, the use of an input-dependent acceptable uncertainty threshold is most beneficial for DDCs with higher dispersion. On the other hand, DDC-based predictions can become useless in some situations, e.g., if there are strict thresholds for the acceptable uncertainty and at the same time their predictions have high dispersion.

## 7    Summary and Conclusion

We proposed novel safety patterns for handling runtime estimations of uncertainty, as they can be provided, for example, by uncertainty wrappers [1]. The patterns support the usage of DDCs to estimate safety-relevant information about the current situation instead of working with static worst-case assumptions.

We observed that the consideration of context information of the driving situation makes it possible to accept more or less uncertainty depending on the inherent risk of the situation. This can lead to a gain in utility, e.g., a reduction of the necessary distance between two vehicles in vehicle platooning. However, the utility gain depends on the concrete application of the patterns. Even for the concrete example of vehicle platooning and DDC-based friction estimation, it is hardly possible to predict the gain in utility manually. For this reason, we developed a tool that allowed us to perform some utility analyses and to quantify the utility gain for the platooning example depending on assumed parameters such as the assumed threshold for acceptable uncertainty. By this means, we were able to answer under which main assumptions the application of the patterns is reasonable.

We conclude that such analyses can provide essential support for early design decisions in the design of increasingly autonomous systems in complex environments because we believe that runtime estimation and handling of uncertainties is necessary to overcome worst-case approximations that would lead to unacceptable utility/performance, especially if the situation context indicates a low risk situation. For this reason, we see the patterns as part of a promising solution to solve this huge challenge by relating the probabilistic target values for safety-relevant functions to uncertainties of DDCs and the confidence that uncertainties are not underestimated.

An open issue in this regard is the consideration of stochastic dependencies between uncertainties estimated in different time steps. Further open issues concern the stochastic dependencies between uncertainties of different DDCs.

## References

[1]     M. Kläs and L. Sembach, "Uncertainty wrappers for data-driven models – Increase the transparency of AI/ML-based models through enrichment with dependable situation-aware uncertainty estimates," in *2nd Int. Workshop on Artificial Intelligence Safety Engineering (WAISE 2019)*, Turku, Finland, 2019.

[2]     M. Kläs and L. Jöckel, "A Framework for Building Uncertainty Wrappers for AI/ML-based Data-Driven Components," in *3rd International Workshop on Artificial Intelligence Safety Engineering (WAISE)*, 2020.

[3]     L. Jöckel and M. Kläs, "Could We Relieve AI/ML Models of the Responsibility of Providing Dependable Uncertainty Estimates? A Study on Outside-Model Uncertainty Estimates," in *40th Int. Conference on Computer Safety, Reliability and Security, SafeComp 2021*, York, United Kingdom, 2021.

[4]     P. Gerber, L. Jöckel and M. Kläs, "A Study on Mitigating Hard Boundaries of Decision-Tree-based Uncertainty Estimates for AI Models," in *Safe AI @ AAAI2022*, Virtual, 2022.

[5]     M. Kläs, R. Adler, I. Sorokos, L. Jöckel and J. Reich, "Handling Uncertainties of Data-Driven Models in Compliance with Safety Constraints for Autonomous Behaviour," in *European Dependable Computing Conference (EDDC)*, 2021.

[6]     F. Arnez, H. Espinoza, A. Radermacher and F. Terrier, "A Comparison of Uncertainty Estimation Approaches in Deep Learning Components for Autonomous Vehicle Applications," in *Workshop in Artificial Intelligence Safety (AISafety)*, 2020.

[7]     B. Kaiser, et al., "Advances in component fault trees," in *ESREL*, 2018.

[8]     S. Kabir, I. Sorokos, K. Aslansefat, Y. Papadopoulos, Y. Gheraibia, J. Reich, M. Saimler and R. Wei, "A runtime safety analysis concept for open adaptive systems," in *6th International Symposium on Model-Based Safety and Assessment*, Thessaloniki, Greece, 2019.

[9]     R. Salay, K. Czarnecki, M. Elli, I. Alvarez, S. Sedwards and J. Weast, "PURSS: Towards Perceptual Uncertainty Aware Responsibility Sensitive Safety with ML," in *SafeAI @ AAAI2020*, New York, 2020.

[10]    M. Henne, A. Schwaiger, K. Roscher and G. Weiß, "Benchmarking Uncertainty Estimation Methods for Deep Learning with Safety-Related Metrics," in *Proceedings of the Workshop on Artificial Intelligence Safety, co-located with 34th AAAI Conference on Artificial Intelligence, SafeAI@AAAI 2020*, New York, USA, 2020.

[11]    J. Reich and M. Trapp, "SINADRA: Towards a Framework for Assurable Situation-Aware Dynamic Risk Assessment of Autonomous Vehicles," in *16th European Dependable Computing Conference (EDCC)*, Munich, Germany, 2020.

[12]    A. Adee, R. Gansch and P. Liggesmeyer, "Systematic Modeling Approach for Environmental Perception Limitations in Automated Driving," in *17th European Dependable Computing Conference (EDCC)*, Munich, Germany, 2021.

[13]    S. Shalev-Shwartz, S. Shammah and A. Shashua, "On a formal model of safe and scalable self-driving cars," *arXiv preprint,* 2017.

[14]    B. Hartmann and A. Eckert, "Road condition observer as a new part of active driving safety," *ATZelektronik worldwide,* vol. 12(5), pp. 34-37, 2017.

[15]    "Predictive road condition services," Robert Bosch GmbH, 2022. [Online]. Available: https://www.bosch-mobility-solutions.com/en/solutions/automated-driving/predictive-road-condition-services/. [Accessed 22 02 2022].

[16]    B. Wassertheurer, Reifenmodellierung für die Fahrdynamiksimulation auf Schnee, Eis und nasser Fahrbahn, Karlsruhe, Germany: KIT Scientific Publishing, 2020.

[17]    G. Van Rossum and F. Drake Jr., Python 3 Reference Manual, Scotts Valley, CA: CreateSpace, 2009.

[18]    T. Kluyver, B. Ragan-Kelley, F. Pérez, B. Granger, M. Bussonnier, J. Frederic, K. Kelley, J. B. Hamrick, J. Grout, S. Corlay, P. Ivanov, D. Avila, S. Abdalla and C. Willing, "Jupyter Notebooks-a publishing format for reproducible computational workflows," in *20th International Conference on Electronic Publishing*, Göttingen, Germany, 2016.

[19]    J. D. Hunter, "Matplotlib: A 2D graphics environment," *Computing in science & engineering,* vol. 9(03), pp. 90-95, 2007.

[20]    C. R. Harris, K. J. Millman, S. J. v. d. Walt, R. Gommers, P. Virtanen, D. Cournapeau, E. Wieser, J. Taylor, S. Berg, N. J. Smith, R. Kern, M. Picus, S. Hoyer, M. H. v. Kerkwijk and M. Brett, "Array programming with NumPy," *Nature,* vol. 585(7825), pp. 357-362, 2020.

[21]    P. Virtanen, R. Gommers, T. E. Oliphant, M. Haberland, T. Reddy, D. Cournapeau, E. Burovski, P. Peterson, W. Weckesser, J. Bright, S. J. v. d. Walt, M. Brett, J. Wilson, K. J. Millman and N. Mayorov, "SciPy 1.0: fundamental algorithms for scientific computing in Python," *Nature methods,* vol. 17(3), pp. 261-272, 2020.